\documentclass[sigconf]{acmart}
\AtBeginDocument{%
  \providecommand\BibTeX{{%
    \normalfont B\kern-0.5em{\scshape i\kern-0.25em b}\kern-0.8em\TeX}}}

\copyrightyear{2024}
\acmYear{2024}
\setcopyright{acmlicensed}
\acmConference[KDD '24] {Proceedings of the 30th ACM SIGKDD Conference on Knowledge Discovery and Data Mining }{August 25--29, 2024}{Barcelona, Spain.}
\acmBooktitle{Proceedings of the 30th ACM SIGKDD Conference on Knowledge Discovery and Data Mining (KDD '24), August 25--29, 2024, Barcelona, Spain}
\acmISBN{979-8-4007-0490-1/24/08}
\acmDOI{10.1145/XXXXXX.XXXXXX}

\usepackage{amsmath}
\usepackage{multicol,multirow}
\usepackage{paralist}
\usepackage{enumitem}
\usepackage{balance}
\setitemize{noitemsep,topsep=0pt,parsep=0pt,partopsep=0pt}
\begin{document}

\title{GRAM: Generative Retrieval Augmented Matching of Data Schemas in the Context of Data Security}

\author{Xuanqing Liu$^\star$, Luyang Kong$^{\star\clubsuit}$, Runhui Wang, Patrick Song,\\ Austin Nevins, Henrik Johnson, Nimish Amlathe, Davor Golac}
\email{{xuanqing, luyankon, runhuiw, patsong, nevinsan, mauritz, amlathe, dgolac} @ amazon.com}
%\email{webmaster@amazon.com}
\affiliation{%
  \institution{Amazon Web Services}
  %\streetaddress{P.O. Box 1212}
  \city{Seattle}
  \state{Washington}
  \country{USA}
}

\renewcommand{\shortauthors}{Liu and Wang, et al.}

\newcommand\blfootnote[1]{%
  \begingroup
  \renewcommand\thefootnote{}\footnote{#1}%
  \addtocounter{footnote}{-1}%
  \endgroup
}
\begin{abstract}
Schema matching constitutes a pivotal phase in the data ingestion process for contemporary database systems. Its objective is to discern pairwise similarities between two sets of attributes, each associated with a distinct data table. This challenge emerges at the initial stages of data analytics, such as when incorporating a third-party table into existing databases to inform business insights. Given its significance in the realm of database systems, schema matching has been under investigation since the 2000s. This study revisits this foundational problem within the context of large language models. Adhering to increasingly stringent data security policies, our focus lies on the zero-shot and few-shot scenarios: the model should analyze only a minimal amount of customer data to execute the matching task, contrasting with the conventional approach of scrutinizing the entire data table. We emphasize that the zero-shot or few-shot assumption is imperative to safeguard the identity and privacy of customer data, even at the potential cost of accuracy. The capability to accurately match attributes under such stringent requirements distinguishes our work from previous literature in this domain.\blfootnote{$^\star$ First two authors contributed equally; $^\clubsuit$ corresponding.}
\end{abstract}




\received{20 February 2007}
\received[revised]{12 March 2009}
\received[accepted]{5 June 2009}

\maketitle

\section{Introduction}
Today's SaaS providers that supports diverse data suppliers with ingesting, managing, and searching for potentially sensitive records, they face the challenge of dealing with inhomogeneous data schemas that naturally occur among different suppliers. For instance, an insurance company may have a table named \textsf{BusinessProfile} that contains columns like \texttt{num\_employees}, \texttt{mailing\_address\_city}, \texttt{business\_phone}, and so forth. However, when we ingest data records for this customer, we discover that the schema is not perfectly aligned with our internal system, which necessitates manually creating mappings such as \texttt{\#employees} $\mapsto $ \texttt{num\_employees}, \texttt{recipient\_city} $\mapsto$ \texttt{mailing\_address\_city} and \texttt{phone\#} $\mapsto$ \\ \texttt{business\_phone}. Unfortunately, these mappings are hardly reusable for another customer due to naming conventions and other nuances. A common scenario in data science domain is that experts spend up to several weeks to designing mappings for moderately sized tables between 100 to 200 attributes. Even with tools like Microsoft BizTalk~\cite{BizTalk} or COMA 3.0 GUI~\cite{COMA3Web}, the data internalization process is typically error-prone and requires multiple rounds of trials and errors.
\par
\begin{figure*}
    \centering
    \includegraphics[width=0.96\linewidth]{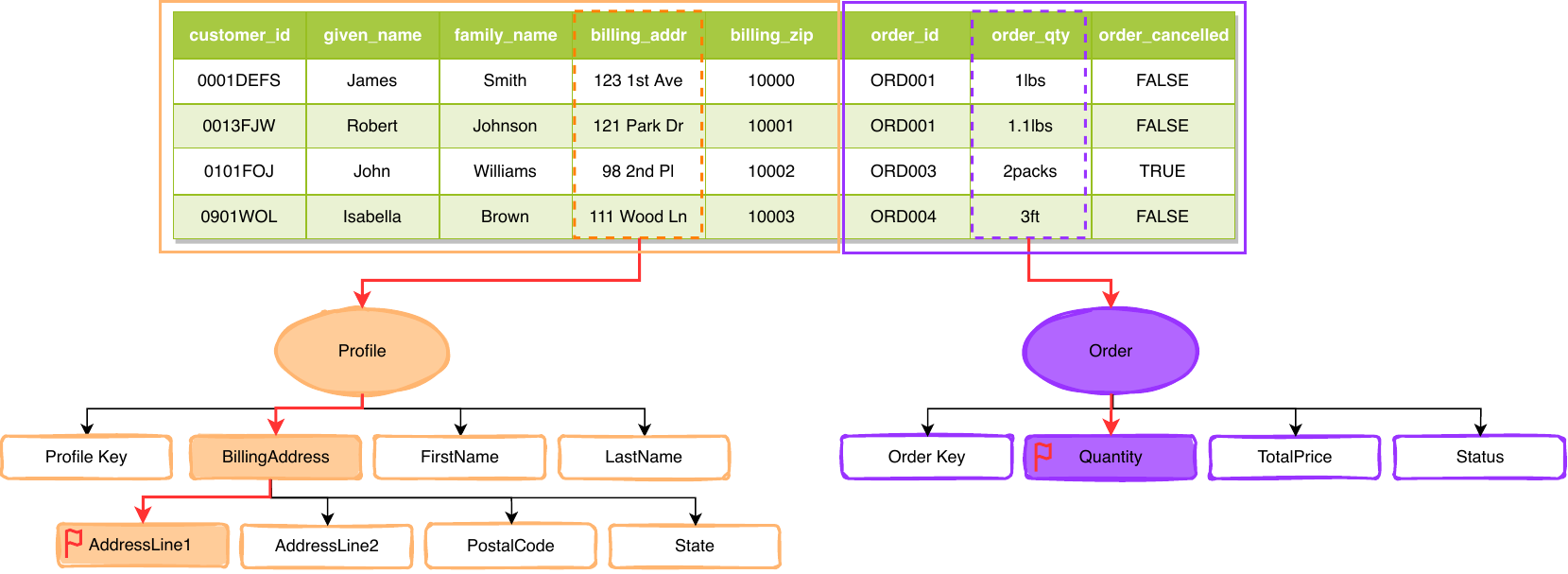}
    \caption{Illustration of the idea of hierarchical prediction in schema mapping. First, columns of input data table are partitioned and grouped into one or more \emph{object types}, here are \texttt{Profile} and \texttt{Order} (two ellipse shapes in figure). Next, we take a column from partition group, then the column traverses through the $n$-ary tree based on the classification results at each level, until a root node is found (marked in red arrows). Each root node corresponds to a target attribute defined by target schema. We repeat the same process for each column until all columns are mapped to target attributes.}
    \label{fig:schema-mapping-example}
\end{figure*}
\begin{figure}
    \centering
    \includegraphics[width=0.9\linewidth]{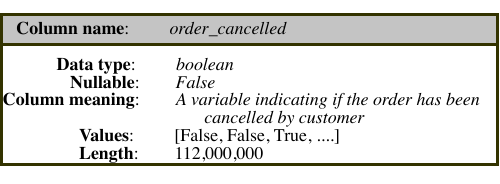}
    \caption{An example of how an individual attribute in the schema look like. We highlight the required field (column name) with shades, and all other fields (data type, nullable, column meaning, values, length, etc.) as optional.}
    \label{fig:schema-mapping-element-example}
\end{figure}
To address the challenge of data ingestion, the research community has proposed the concept of automated schema matching as a solution to streamline the associated processes. This concept is visually represented in Figure~\ref{fig:schema-mapping-example}. In essence, it transforms the attribute-to-attribute mapping task into a hierarchical multi-class classification problem. Given an input table with $N$ columns, the approach involves a non-overlapping partition of the $N$ columns into $k$ subgroups, denoted as $N_1, N_2, \dots, N_k$ columns, where $\sum_{i=1}^k N_i=N$. Each subgroup corresponds to a distinct object type, exemplified in Figure~\ref{fig:schema-mapping-example} by showcasing the \texttt{Profile} and \texttt{Order} object types. For each object type, a predefined attribute tree is established, comprising nodes and attributes (with leaf nodes serving as aliases for attributes). By deploying a classifier at each non-leaf node to predict the correct child node containing the relevant attribute, the methodology simplifies the process to traversing an $n$-nary tree. This traversal follows the direction indicated by classification results at each level. As discussed later, numerous prior works align with and contribute to this overarching framework.
\par
Diverging from prior research efforts, we reexamine the aforementioned issue by harnessing the latest advancements in language understanding, specifically leveraging Large Language Models (LLMs) and their adept in-context learning capabilities. By encapsulating the previously outlined hierarchical classification problem within the framework of in-context learning, we proficiently repurpose LLMs as readily available classifiers through the mechanism of few-shot prompting.
\par
Our primary contributions can be summarized as follows:
\begin{compactenum}
\item \noindent We address the automated schema matching problem within the context of data privacy, employing a novel perspective that incorporates zero-shot and few-shot predictions.
    \item Our solution seamlessly integrates the recent surge in Large Language Models (LLMs). We conduct a comprehensive benchmarking exercise across various open-source and proprietary LLMs to assess their performance.
    \item Introducing a dynamic prompt selection method based on input characteristics, our approach not only enhances inference speed but also augments the in-context learning accuracy of LLMs.
    \item Beyond the conventional scope of schema matching, our solution incorporates object type detection and unique key detection. These additional components transform the standalone schema matching module into a more feature-complete data-table ingestion service.
    \item We rigorously benchmark the accuracy of our methodology against relevant approaches using both public and production-quality, synthetic datasets. Particularly noteworthy is the utilization of datasets designed to mirror realistic workloads in various industrial applications.
\end{compactenum}
\section{A Brief History of Schema Matching Research}
\subsection{Pioneering solutions}
\paragraph{LSD}\cite{doan2000learning} stands as one of the pioneering machine learning-based schema matching frameworks. It formulates the matching problem as a multi-class classification challenge. Notably, LSD employs an ensemble of classifiers to enhance accuracy, incorporating a nearest neighbor Whirl learner, a Naive Bayesian learner, a name matcher, and a county-name recognizer. Classifiable under the dichotomies outlined earlier, LSD falls within the category of one-to-one matching based on linguistic features and is trained on both schema and instances.
\paragraph{CUPID} \cite{madhavan2001generic} is considered one of the first general-purpose schema matching systems with a focus on feature completeness. It employs linguistic features and predefined rules to match pairs or groups of attributes. CUPID's core idea is to determine the highest weighted similarity ($wsim$) between two attributes using the formula $wsim = w_{\mathrm{struct}} \cdot ssim + (1 - w_{\mathrm{struct}}) \cdot lsim$, where $ssim$ is the structural similarity score, and $lsim$ is the linguistic similarity score. As an early work from the 2000s, CUPID's feature extractors are basic compared to modern language models. However, CUPID falls short in extracting insights from column values, missing opportunities to address ambiguities inherent to schema-data alone.

\paragraph{Similarity Flooding}~\cite{melnik2002similarity} introduces a method to transform the schema matching problem into computing the fixpoint over graph propagation. Initially, the SQL2Graph operation converts a pair of table schemas into two graphs for matching. The StringMatch operation assigns initial similarity scores over nodes in the graphs. Subsequently, the SFJoin operation, essentially a label propagation algorithm over a directed graph, is applied to iteratively obtain the fixpoint. Attribute pairs are then pruned based on a specified threshold. Similar to CUPID, the text similarity metric appears basic by contemporary standards, considering only the length of common prefixes and suffixes between two strings. Additionally, it does not incorporate column values, rendering it suboptimal for challenging use cases.
\paragraph{COMA/COMA++/COMA 3.0} \cite{do2002coma,aumueller2005schema,massmann2011evolution} constitute a line of research that focuses on \underline{co}mbining \underline{ma}tching algorithms in a flexible manner, thus presenting an orthogonal approach to the methods discussed earlier. The notable aspect of this software system, along with the underlying algorithms, is its provision of a user-friendly interface for executing multiple matching algorithms iteratively, allowing for human intervention. Additionally, the software extends beyond merely matching two schemas, encompassing a comprehensive workflow that includes storage, match execution, mapping processing, and user connectivity.

\paragraph{S-MATCH} \cite{giunchiglia2004s} shares similarities with the COMA family as it is an open-source framework that provides multiple built-in matching algorithms. Users can readily adopt the system and make necessary extensions as required.

\subsection{Modern solutions based on neural nets}
\paragraph{Sahay \emph{et al.}} \cite{sahay2020schema} presented a straightforward hybrid approach incorporating both schema data and column values, applicable to both one-to-one and one-to-many mappings. Employing extensive feature engineering, the authors utilize self-organizing maps or K-means clustering to cluster similar attributes. Consequently, during testing, an attribute is paired with the nearest cluster, and the best attribute within that cluster is selected based on the minimum edit-distance principle.

\paragraph{H\"attasch \emph{et al.}} \cite{hattasch2022s} introduced a neural embedding-based method, making a significant contribution with a two-level matching scheme. The first level involves table matching, followed by attribute matching at the second level. The matching process entails computing the similarity, such as cosine similarity, between two embeddings derived from textual information, including column name, data type, comments, etc.

\paragraph{LSM} \cite{zhang2023schema} is a schema matcher that leverages pre-trained language models. Notably relevant due to its recent development and utilization of modern transformer-based language models, LSM employs a finetuned BERT featurizer at its core. This featurizer transforms pairs of schema information into similarity scores, considering two attributes as a match if the similarity score surpasses a specified threshold. The BERT featurizer undergoes finetuning based on human-provided labels. Once the finetuning process is complete, the model is prepared to generalize to new schema pairs.

\subsection{Goals of our solution}
Given the recent surge in large language models (LLM) and generative AI (GenAI), it is intuitive to explore the application of the "emergent abilities" described by Wei et al.~\cite{wei2022emergent} to the realm of schema matching. Our decision to integrate these advancements stems from the belief that LLMs offer language understanding and reasoning capabilities approaching human levels. With this upgrade, we anticipate a transformative impact on how we conceptualize the similarity between two data schemas. In this paper, we aim to elevate the quality and usability of schema matching systems along the following dimensions:

\paragraph{Enhanced Language Understanding with Efficient Inference.} When framing the schema mapping as a natural language processing (NLP) problem, one observes that the advancements in solutions reviewed over the past two decades are intricately linked to the evolving landscape of language modeling. Early solutions relied on string similarity and hand-crafted features, often complemented by shadow models such as linear classifiers, naive Bayes classifiers, or $k$-means clustering methods. In contrast, contemporary solutions leverage deep learning text featurizers like Word2vec, GloVe, FastText, and BERT, extracting text similarity scores within an end-to-end paradigm. This paper benefits from superior language understanding capabilities offered by open-source large language models, specifically the FLAN-T5 family. Additionally, we introduce methods to expedite inference speed while preserving accuracy, a critical consideration for handling large-scale data prevalent in industrial applications.

\paragraph{Minimal Training Data Dependency.} The conventional approach to utilizing finetuned language models, as seen in works such as \cite{zhang2023schema}, involves the collection of a substantial amount (ranging from $10^3$ to $10^4$) of human-labeled data to calibrate the language classifier using certain loss functions. In contrast, we adopt zero-shot and few-shot learning, also known as in-context learning (ICL) in Large Language Models (LLM) literature, reducing the dependency on data quantity. This attribute is particularly significant in addressing contemporary concerns regarding data security and privacy, as it obviates the need for accessing and annotating large volumes of customer data.

\paragraph{Comprehensive Feature Integration.} The solution presented in this paper transcends the boundaries of a mere schema matcher, evolving into an end-to-end service fueled by language models. This service harmonizes disparate data sources, rendering them into uniformly searchable data records. Central to this endeavor are two supplementary components around the attribute mapper: the object type detector and column key detector. Both components leverage language models to enhance their functionalities. Specifically, the object type detector identifies the appropriate object type (target schema) for a subset of input columns; the attribute mapper establishes connections between each input attribute and a unique target attribute; and finally, the key detector designates one of the attributes as the unique key, enabling the ingestion of the entire table with duplicates removed.

\section{Background Knowledge}
\subsection{Large language model}
Language understanding stands as a fundamental capability to showcase advanced artificial intelligence. Pretrained language models (PLM)~\cite{devlin2018bert} have proven to be a powerful and scalable approach for embedding general knowledge into transformer-based neural networks. The conventional application of PLMs involves finetuning them on domain-specific datasets collected from experts, leading to the creation of one model for each task. This practice, however, limits usability in scenarios where high-quality datasets are scarce. With the growing demand for more generalized language models, researchers have identified a promising avenue. By scaling up both the size of the pretraining corpus and the parameter count of the language model, adhering to scaling-up principles~\cite{kaplan2020scaling,tay2021scale}, and subsequently finetuning the model on diverse tasks using instructional prompts~\cite{chung2022scaling,longpre2023flan,ouyang2022training}, a robust language model emerges. This model exhibits the capability to comprehend natural instructions with strategic prompt engineering.

\subsection{Retrieval augmentation\label{sec:RAG}}
Standalone Large Language Models (LLMs) encode world knowledge within their model weights, placing smaller scale models at a disadvantage when tasked with memorizing intricate language corpora that demand hundreds of billions of parameters. Additionally, Retrieval Augmented Generation (RAG)~\cite{lewis2020retrieval} enhances model capacity by integrating an external knowledge search engine. RAG excels in consolidating domain-specific knowledge that proves challenging to memorize from a web corpus using LLM-based readers. In this context, RAG emerges as particularly well-suited for the schema matching task, given the often vaguely defined connections between two attributes.

\section{GRAM: Generative Retrieval Augmented Matching}
\subsection{Motivating example}
\begin{figure}[htb]
    \centering
    \includegraphics[width=1.0\linewidth]{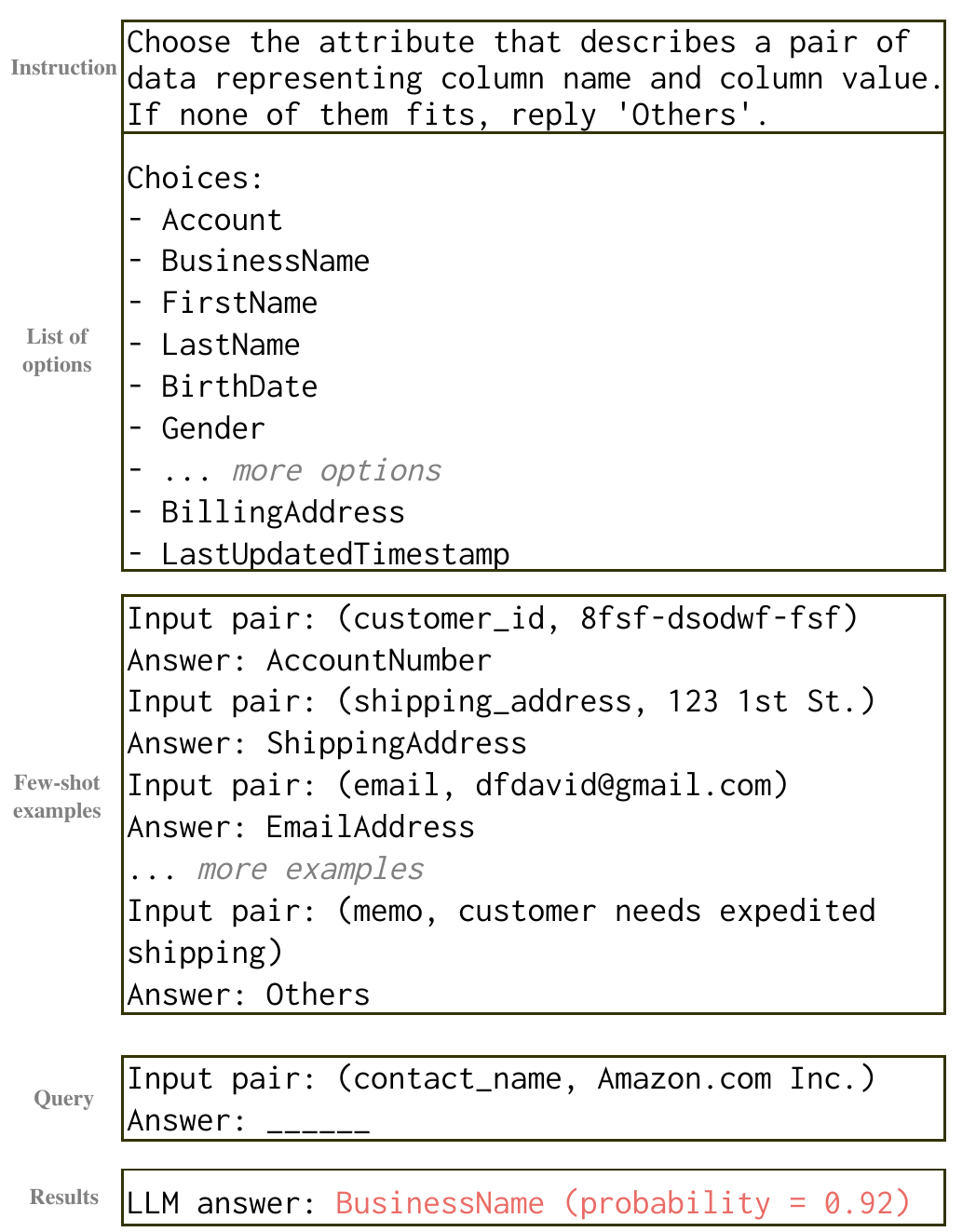}
    \caption{An illustrative example outlining the concept of prompting Large Language Models (LLMs) to match a source attribute (e.g., contact\_name for Amazon.com Inc.) to a list of $15$ target attributes is provided for clarity.}
    \label{fig:intuitive-example}
\end{figure}

To explore how instruction finetuned large language models can be prompted with few-shot examples to effectively match similar attribute pairs against unrelated ones, we have developed a straightforward demonstration using Anthropic Claude via the AWS Bedrock SDK~\cite{bedrock}, as illustrated in Fig.~\ref{fig:intuitive-example}.
In this instance, we instruct Claude to match the column name \texttt{contact\_name} with an example value \texttt{Amazon.com Inc.} against other profile-related attributes, such as \texttt{FirstName}, \texttt{LastName}, \texttt{HomePhoneNumber}, \\\texttt{EmailAddress}, and so on. The prompt adheres to the standard in-context learning paradigm: it begins with a formulation of the problem statement and the success goal, followed by a list of choices and subsequently a list of examples with ground-truth labels. Finally, the query example is appended at the end.
\par
It is noteworthy that this particular problem is non-trivial to answer accurately. The column name \texttt{contact\_name} alone can refer to both \texttt{FirstName}, \texttt{LastName}, and \texttt{BusinessName}. The resolution of this ambiguity is dependent on examining the example value \texttt{Amazon.com Inc.}, where the model deduces that \texttt{BusinessName} is the sole appropriate match. Generally, the schema matching problem proves to be highly challenging, even for domain experts. For instance, the column name \texttt{state} may represent a U.S. state name or serve as an equivalent to the word ``\texttt{status},'' without additional information discernible from the value section.
\par
We hypothesize that leveraging large language models equipped with commonsense knowledge represents a promising approach to effectively address the challenge of schema understanding. Concurrent research indicates that gigantic language models boasting 100+ billion parameters demonstrate human-level reading comprehension and near-human-level logical reasoning capabilities~\cite{wei2022emergent}. This hypothesis serves as the driving force behind our decision to incorporate an instruction-finetuned large language model as the central component of our schema matching service.
\par
However, translating the concept illustrated in Fig.~\ref{fig:intuitive-example} into live production proves to be non-trivial. The target processing speed of schema matching service is  $10$ transactions per second (TPS) per host, each equipped with inference-optimized GPU devices, typically Nvidia T4 or Nvidia A10. Benchmark results reveal that, without any optimization, the naive solution achieves less than $6$ TPS per host.
\par
In the subsequent sections, we delve into strategies for accelerating inference latency, or equivalently, increasing the TPS count. While various techniques exist for optimizing Large Language Model (LLM) inference, including intelligent decoding methods~\cite{leviathan2023fast,santilli2023accelerating}, improved memory access patterns~\cite{kwon2023efficient,dao2022flashattention}, and model compression and quantization~\cite{frantar2022gptq,xiao2023smoothquant}, among others, this paper introduces an orthogonal approach specifically tailored for schema matching acceleration, known as \emph{prompt compression}. Our approach is inspired by the observation that the inference time $T$ for transformer-based LLMs is quadratic to the input length $L_{\mathrm{input}}$, i.e., $T=\mathcal{O}(L_{\mathrm{input}}^2)$. This is because the self-attention output is computed as
\begin{equation}\label{eq:self-attention}
X_{\mathrm{out}}=Softmax\Big(\frac{Q^TK}{\sqrt{d}}\Big)V,
\end{equation}
where $Q=W_q^TX_{\mathrm{in}}$, $K=W_k^TX_{\mathrm{in}}$, $V=W_v^TX_{\mathrm{in}}\in\mathbb{R}^{d\times L_{\mathrm{input}}}$ are attention query, key and value matrices respectively; $X_{\mathrm{in}}$ and $X_{\mathrm{out}}$ are the inputs and outputs of attention block. The bottleneck for computing Eq.~\eqref{eq:self-attention} is matrix multiplication $Q^TK$ with a complexity of $\mathcal{O}\big(L_{\mathrm{input}}^2d\big)$. As a result, it is most beneficial to minimize the input text length $L_{\mathrm{input}}$ to accelerate the inference speed. At the same time, according to the prompt structure in Fig.~\ref{fig:intuitive-example}, we can decompose $L_{\mathrm{input}}$ to
\begin{equation}\label{eq:input-length-decompose}
L_{\mathrm{input}}=L_{\mathrm{instruct}}+N\cdot\big(\bar{L}_{\mathrm{option}}+M\cdot\bar{L}_{\mathrm{example}}\big),
\end{equation}
where $L_{\mathrm{instruct}}$ is the length of instruction text, $\bar{L}_{\mathrm{option}}$ is the average length of destination attribute name, $\bar{L}_{\mathrm{example}}$ is the average length of each example; $N$ is the number of options in prompt, and this is equivalent to number of mapping destinations; $M$ denotes number of examples per option ($M$-shot prompting).
\par
Our empirical observation indicates that listing all possible matching destinations in each Large Language Model (LLM) query is unnecessary. Instead, by employing a combination of techniques detailed in the following sections, we can effectively eliminate a substantial number of irrelevant options and examples associated with the source attribute. This results in a significant reduction in the values of $N$ and $M$ in Eq.~\eqref{eq:input-length-decompose}. Consequently, a smaller value for $L_{\mathrm{input}}$ is achieved.

\subsection{NER-based destination filter\label{sec:NER}}
Named Entity Resolution (NER) serves as a potent method for extracting and recognizing categorical information from free texts. For example, consider the text:
\begin{quote}
    ``Jim bought 300 shares of Acme Corp. in 2006.''
\end{quote}
A successful NER task would label ``Jim'' as \textbf{Person}, ``Acme Corp.'' as \textbf{Organization}, and ``2006'' as \textbf{Time}. With NER models, we can move beyond merely matching the column data type, as seen in prior works (e.g., \cite{aumueller2005schema}), to introduce a new destination attribute filter denoted as $\mathcal{F}_{\mathrm{NER}}$. This filter retains only those destination attributes that share both the same data type and data category as the source attribute. Mathematically,
\begin{equation}\label{eq:NER-filter}
\begin{split}
\mathcal{F}_{\mathrm{NER}}\big(S|\langle k,v\rangle\big)=\big\{o | o\in S &\wedge \mathrm{DType}(o)=\mathrm{DType}(v) \\
&\wedge \mathrm{NER}(o)=\mathrm{NER}(v)\big\},
\end{split}
\end{equation}
in which $S$ is the set of all destination attributes; $\langle k,v\rangle$ is the input data pair containing attribute name $k$ and attribute value $v$; $\mathrm{DType}$ is the data type extraction operator by reading column metadata; $\mathrm{NER}$ denotes a named entity resolution model we trained on schema matching tasks.
\par
To highlight the potential impact of the filter $\mathcal{F}_{\mathrm{NER}}$, let's revisit the prompt depicted in Fig.~\ref{fig:intuitive-example}. Post-filtering, the available options are significantly reduced to just two - \texttt{Account} and \texttt{BusinessName}, down from the original 15 options. This reduction is attributed to the NER model's recognition of the input value \texttt{Amazon.com Inc.} as an organization name, while the remaining options fall into distinct data categories such as phone numbers, person names, addresses, etc.
\par
We implemented a Named Entity Recognition (NER) model tailored for schema matching tasks, closely adhering to standard practices outlined in the literature (\cite{li2020survey} and references therein), with a few noteworthy modifications. First, we defined a more fine-grained label space. Traditional NER models are typically trained on a coarser label space, where the target category "address" encompasses street addresses, cities, states/provinces, and even countries. However, this standard practice limits usability in schema matching tasks where the goal is to determine if a column storing zip codes matches another column storing cities, even if both are mapped to the "address" category with traditional NER models. The second modification we introduced pertains to the training loss. In traditional NER models, the loss is computed on a per-token basis, treating it as a token-level classification problem. This approach is justified when the input is a sentence containing multiple entities, and the goal is to predict the text span encompassing all entities along with their labels. In contrast, our approach computes the loss at the sequence level, treating it as a sequence-level classification problem. Our approach is valid under the assumption that there is only one entity for each input sequence, a condition that is widely applicable to schema matching tasks.
\par
Implementation-wise, we choose RoBERTa-base~\cite{liu2019roberta} as the backbone model to initialize training. An input sequence for training or inference consists of a few samples ranging from 1 to 6 elements sampled from a column, serialized as a list of values
\begin{equation}\label{eq:NER-data-serialize}
\langle s\rangle\{\mathrm{value}_1\}[SEP]\{\mathrm{value}_2\}[SEP]\cdots\{\mathrm{value}_k\}\langle/s\rangle, 1\le k\le 6,
\end{equation}
$\langle s\rangle, \langle/s\rangle, [SEP]$ are special tokens in vocabulary, $\{\mathrm{value}_i\}$ is the $i$-th sample of the column, and To enhance robustness and generalizability, we employ random sampling, selecting $1 \leq k \leq 6$ examples to construct a training sequence. For additional training details, please refer to the appendices.

\subsection{Double-RAG filter\label{sec:double-RAG}}
The NER-based filter discussed in the previous section assesses the coherence of two attributes based on column values. Essentially, two attributes can be considered a good match when their corresponding column values are mapped to the same Named Entity Recognition (NER) label. In this section, we adopt a different perspective and gauge the inter-attribute coherency through the semantic similarity of column names. Our approach draws inspiration from the efficacy of the retrieval augmentation (RAG) technique in enhancing accuracy across various Large Language Model (LLM) applications (refer back to Section~\ref{sec:RAG} for additional background). What sets our use of the RAG technique apart is that we not only search for the best possible options but also seek the most suitable few-shot examples for a particular option, giving rise to the term "Double-RAG." Consequently, both the options and the few-shot examples in the prompt dynamically change with different input queries. In this regard, our proposed prompting method can be viewed as another instance of automatic prompt tuning~\cite{zhou2022large}, with the goal of minimizing the prompt length while maintaining robust reasoning abilities.
\par
We maintain two databases to store the options and examples for each option. Let $D_{\mathrm{opt}}=\{o_1, o_2, \dots, o_N\}$ be the database containing options (destination attribute names) and
$$
D_{\mathrm{ex}}=
\begin{Bmatrix}
e_{11} & e_{12} & \cdots & e_{1M} \\
e_{21} & e_{22} & \cdots & e_{2M} \\
\vdots & \vdots & \ddots & \vdots \\
e_{N1} & e_{N2} & \cdots & e_{NM}
\end{Bmatrix}
$$
be the database containing all available examples, in which $e_{ij}$ is the $j$-th example of $i$-th option. We provide $M$ examples for each of the $N$ options, totaling $NM$ examples. To understand how members in each database look like, we can pick a few examples from both. Suppose $o_1=$``PhoneNumber'', then we can store $e_{11}=$``phone'', $e_{12}=$``tel'', $e_{13}=$``phone\_number'' and so on. In principle, we should collect a diverse number of examples that give LLM enough idea of how the concept of option $o_i$ is like.
\par
Building upon the databases $D_{\mathrm{opt}}$ and $D_{\mathrm{ex}}$, we further incorporate a similarity measure $\mathrm{sim}(x, y)\in[0, 1]$, supported by either machine learning models or traditional string similarity algorithms. The trade-off in this context revolves around whether the critical factor is the semantic understanding ability from machine learning models and the computational budget available in practice.
\par
For instance, we anticipate the similarity value of $\mathrm{sim}(\text{phone}, \text{tel})$ to be closer to $1.0$, but none of the string similarity algorithms yields expected results in such cases without the aid of external thesaurus dictionaries. This is because the words "phone" and "tel" share only one common character, "e", resulting in a bi-gram Jaccard similarity of $0.0$. In contrast, even the simplest GloVe~\footnote{URL: \url{https://nlp.stanford.edu/projects/glove/}, we used glove.42B.300d.zip version.} embedding indicates a significant cosine similarity of $0.50$, not to mention more sophisticated BERT-based embeddings. The experiments will revisit the choice of similarity measures with further details.
\par
Equipped with two databases $D_{\mathrm{opt}}$ and $D_{\mathrm{ex}}$, and a similarity measure $\mathrm{sim}(x, y)\in[0, 1]$, we are ready to formulate the way we short-listing the options together with their exemplars:
\begin{equation}\label{eq:double-rag}
\begin{aligned}
\widehat{D}_{\mathrm{opt}}&=\Big\{o_i | o_i\in D_{\mathrm{opt}} \wedge i\in\mathrm{Top}_{k_1}\big(\mathrm{sim}(o_i, q)\big)\Big\},\\
\widehat{D}_{\mathrm{ex}}&=\Big\{e_{ij} | e_{ij}\in D_{\mathrm{ex}} \wedge o_i\in\widehat{D}_{\mathrm{opt}} \wedge j\in{\mathrm{Top}_{k_2}\big(\mathrm{sim}(e_{ij}, q)\big)} \Big\}. 
\end{aligned}
\end{equation}
Above we defined $q=\langle k, v\rangle$ as the key-value query pair; $\widehat{D}_{\mathrm{opt}}$ and $\widehat{D}_{\mathrm{ex}}$ as two compressed databases by filtering out the dissimilar choices and exemplars to $k_1\ll N$ and $k_1\cdot k_2\ll NM$ elements, respectively. 
\subsection{Other components}
For the sake of comprehensive service functionality, we have designed two additional components that work in conjunction with the core Large Language Model (LLM)-based attribute mapper to execute the data integration task. These components are the object type detector and key detector. While the primary focus of this paper revolves around attribute mapping, as an integral part of the overall system, we briefly introduce their functionalities as follows.
\par
\paragraph{Object Type Detector} This component serves as a preprocessor for the attribute mapper. Its role is to partition the columns of the input table into multiple subgroups, each representing a uniform topic (also referred to as an object type, as illustrated in Fig.~\ref{fig:intuitive-example}), such as personal profile, customer order, issue ticket, and so forth. It is important to note that real-life input tables can be a combination of multiple topics, and the two databases $D_{\mathrm{opt}}$/$D_{\mathrm{ex}}$ used in the LLM attribute mapper are determined by the topic. Hence, the system needs to cluster the columns and identify the topic of each cluster before proceeding to the attribute mapping stage. Our implementation of the object type detector adheres to standard practices: we first convert the input table into CSV format, then serialize its header into a text string. Next, the entire string is tokenized to train a BERT-based multi-class classifier with per-token level cross-entropy loss. During the inference stage, we group columns with the same predicted labels together into a subgroup, effectively partitioning the entire table.

\paragraph{Key Detector} This component functions as a postprocessor for the attribute mapper. Its role is to enhance the mapping results with a few keys for searching or de-duplication. The underlying concept aligns with the LLM-based attribute mapper introduced earlier; in fact, we reuse the same LLM model with a different prompting method, thereby improving hardware utilization rates. Initially, we allow users to customize heuristic rules to filter out columns unlikely to serve as potent keys. A simple illustrative rule could be any column name with the pattern ``*\_id''. Users have the flexibility to chain multiple rules together to strike a balance between recall and precision. Ideally, the aim is to retain all valid keys while minimizing the list of candidates to query LLM.

\subsection{Workflow}
\begin{figure*}[htb]
    \centering
    \includegraphics[width=0.99\linewidth]{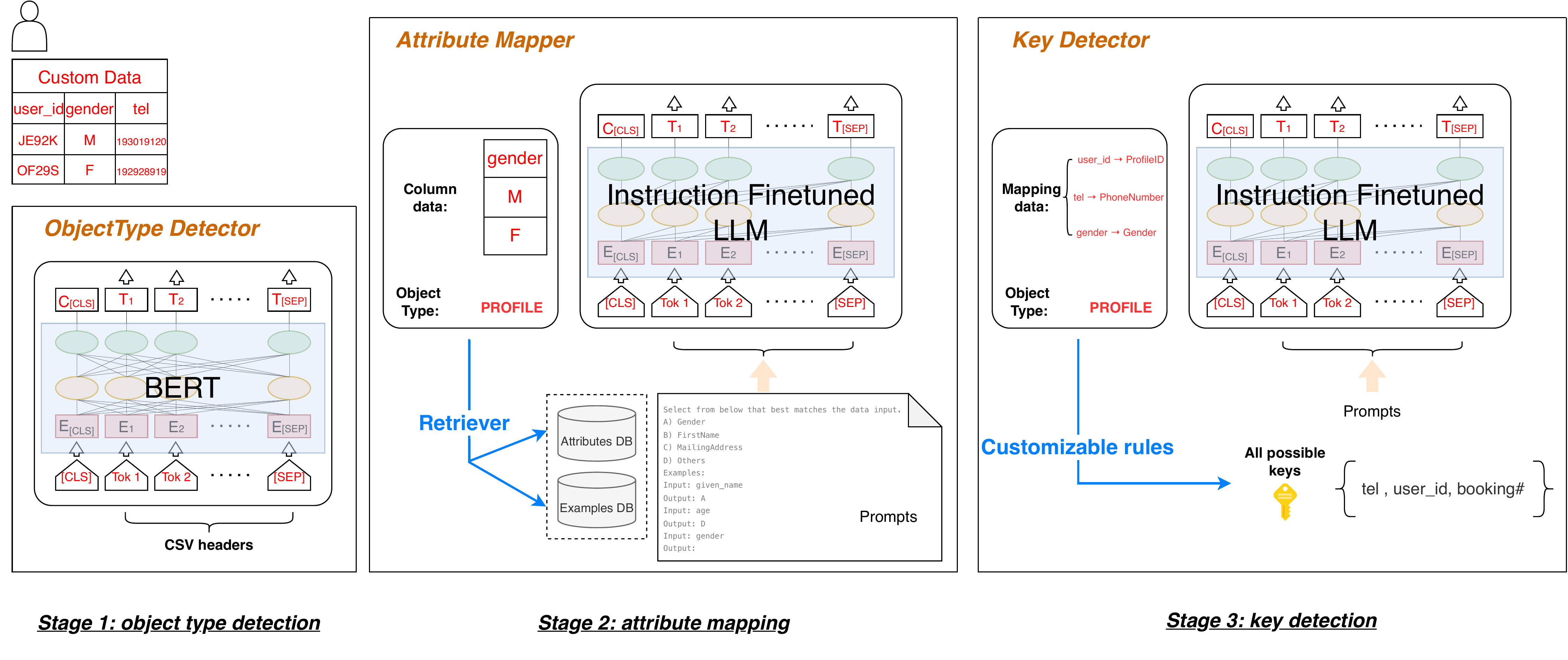}
    \caption{Architecture and workflow of GRAM.}
    \label{fig:architecture}
\end{figure*}
Bringing all the components together, we illustrate the entire workflow in Fig.~\ref{fig:architecture}. At a high level, the custom data slated for ingestion first undergoes the object type detector, where columns are partitioned and labeled with one of the pre-defined object types. In the second stage, each individual column, along with its associated object type, is formatted as a query to the attribute mapper. The outcome of stage 2 is the predicted destination attribute generated from the instruction-finetuned Large Language Model (LLM). Finally, in stage 3, the key detector assigns one or more keys to the mapped attributes, ensuring that the ingested table is accompanied by keys for searching and data de-duplication.

\section{Experiments}
We have designed a series of experiments to assess the effectiveness of the Large Language Model (LLM)-based attribute mapper. Specifically, we aim to address the following questions:
\begin{enumerate}
    \item How does this retrieval-augmented LLM solution compare with traditional solutions in terms of accuracy?
    \item What benefits are observed in throughput when incorporating the prompt compression techniques discussed in Sec.~\ref{sec:NER} and \ref{sec:double-RAG}?
    \item What is the most practical choice among LLM backbones of different sizes?
    \item How does the number of $k$-shot examples influence end-to-end accuracy?
\end{enumerate}

We initiate the process with dataset preparation, which is undeniably one of the most challenging steps given the multi-decade history of schema mapping research since the 2000s. Numerous datasets referenced in early works are either lost or unpublished. Despite these challenges, we have managed to reconstruct a substantial collection of evaluation sets from diverse domains, as listed below.

\begin{itemize}
    \item \textbf{Personal Contacts:} This domain revolves around personal and business profiles, which are commonly found in customer databases, employee databases, or social media records. In total, there are $1400$ columns.
    \item \textbf{Sales:} This domain encompasses sales and transaction records for a merchant, such as airline bookings and shopping checkouts. In total, there are $400$ columns.
    \item \textbf{Products:} This domain comprises databases storing products or services available in the market, including airlines, hotel rooms, groceries, etc. In total, we have collected $200$ columns.
    \item \textbf{Issue Tickets:} This domain includes issue tickets, totaling $330$ columns.
\end{itemize}

\textbf{PII Disclosure:} None of the datasets mentioned above contain any real identity information. This includes metadata such as column names and/or data types (\texttt{first\_name(str)}, \texttt{dob(str)}, \texttt{zip(int32)}, \texttt{address\_line1(str)}, \texttt{sales\_amount(float32)}). The column values are all synthetic or randomly generated.
\par
We have implemented and deployed our Large Language Model (LLM)-based schema matching system using PyTorch~\cite{paszke2019pytorch}, based on the FLAN-T5 model. For very early methods, such as LSD~\cite{doan2000learning} and CUPID~\cite{madhavan2001generic}, for which no first-party implementation is available, we implemented their methods following the ideas presented in the original papers. For other similar works, such as Similarity Flooding~\cite{melnik2002similarity}, we were unable to replicate the algorithm due to the lack of critical details; hence, we exclude them from our experiments. When benchmarking throughput, we executed all programs on hardware comprising $4\times$ Nvidia A10 GPUs (each with 24GB of memory), 24 physical CPU cores, and 192GB of memory.
\subsection{Comparing LLM-based schema matching with baselines}
\begin{table}[htb]
    \centering
    \begin{tabular}{r|ccccc}
    \toprule
    \multirow{2}{*}{Algorithms} &  \multicolumn{5}{c}{Mean accuracy (\%) in domain}\\
    \cmidrule{2-6}
           & Person & Sales & Products & Tickets & Avg.\\
    \midrule
    LSD~\cite{doan2000learning}    & $73.0$ & $63.6$   & $61.8$  &  $74.7$  & $68.3$ \\
    CUPID~\cite{madhavan2001generic} & $52.2$ & $50.6$   & $39.8$   & $62.7$  & $51.3$ \\
    COMA 3.0~\cite{COMA3Web}  & $56.6$ & $48.7$   & $69.0$  &  $50.6$ &  $56.2$ \\
    LSM~\cite{zhang2023schema}    & $81.0$ & $78.5$   & $70.2$  &  $71.4$  & $75.3$ \\\hline
    GRAM (ours) & $\mathbf{91.9}$ & $\mathbf{80.3}$ & $\mathbf{92.3}$ & $\mathbf{90.3}$ & $\mathbf{88.7}$ \\
    \bottomrule
    \end{tabular}
    \caption{Comparing the accuracy numbers across different domains among traditional algorithms, deep neural nets based algorithms, and our LLM based algorithm.}
    \label{tab:main-comparison}
\end{table}
In this study, we conduct a comprehensive comparison of various schema matching algorithms. The primary objective is to assess the comparative advantages of machine learning (ML)-based and large language model (LLM)-based algorithms in comparison to conventional rule-based methods. The outcomes are presented in Figure~\ref{tab:main-comparison}. Based on the experimental findings, several observations can be made: 1) a noteworthy improvement in accuracy is evident when employing an instruction-finetuned LLM, surpassing even contemporary pretrained language model (LM) approaches, such as LSM~\cite{zhang2023schema}; 2) it is generally observed that embedding-based cosine similarity complements lexical similarity metrics effectively. Our internally developed implementation of the LSD method exhibits noteworthy performance, particularly when utilizing an ensemble of word embeddings and the Sorensen-Dice~\cite{sorensen1948method} string similarity algorithm.
\subsection{Effect of NER-based filter and Double-RAG filter}
\begin{table}[htb]
    \centering
    \begin{tabular}{r|ccccc}
    \toprule
    \multirow{2}{*}{Settings} &  \multicolumn{5}{c}{Mean accuracy (\%) in domain}\\
    \cmidrule{2-6}
              & Person & Sales & Products & Tickets & Avg.\\
    \midrule
    No filter     & $83.0$ & $78.7$ & $81.1$ & $90.5$   & $83.3$ \\
    +NER          & $\mathbf{92.7}$ & $\mathbf{86.6}$ & $88.1$ & $85.4$   & $88.2$ \\
    +Double-RAG   & $89.4$ & $74.9$ & $89.6$ & $\mathbf{91.6}$   & $86.4$ \\
    +Both filters & $91.9$ & $80.3$ & $\mathbf{92.3}$ & $90.3$   & $\mathbf{88.7}$ \\ 
    \bottomrule
    \end{tabular}
    \caption{Comparing the testing accuracy with and without filters. Filters do not change model accuracy in a consistent direction.}
    \label{tab:filter-ablation}
\end{table}
We explore the impact of the Named Entity Recognition (NER) filter and the Double-RAG filter on both inference speed and accuracy. In principle, activating either of these filters introduces false negatives, as they possess the capability to exclude positive selections and crucial instances intended to assist reasoning during test time. The discernible effect of these filters on accuracy is detailed in Table~\ref{tab:filter-ablation}.
\begin{figure}[htb]
    \centering
    \includegraphics[width=0.85\linewidth]{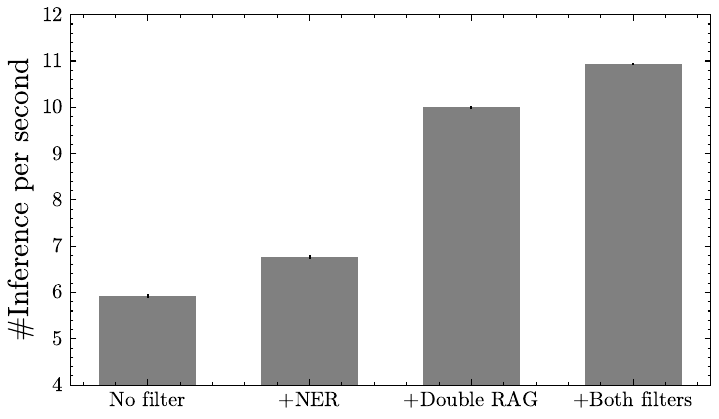}
    \caption{Inference speedup due to NER and double-RAG filters. With double-RAG filter, we keep $k_{\mathrm{opt}}=4$ options and $k_{\mathrm{ex}}=1$ examples for each of the $4$ options. Error bars are provided but barely visible.}
    \label{fig:filter-speedup}
\end{figure}
Remarkably, a consistent decline in accuracy is not readily observable upon activating the filters. We posit that the incorporation of high-quality filters aids the Large Language Model (LLM) in decision-making by eliminating evidently incorrect option items and unrelated few-shot examples. Despite introducing false negatives through occasional removal of correct options and valuable few-shot examples, the overall impact appears to enhance the LLM's decision-making process. Meanwhile, the filters demonstrate a noticeable acceleration in inference speed, as illustrated in Figure~\ref{fig:filter-speedup} across typical workloads simulated with synthetic datasets.

\subsection{Does larger language model perform better in schema-matching?}
\begin{figure}[htb]
    \centering
    \includegraphics[width=0.85\linewidth]{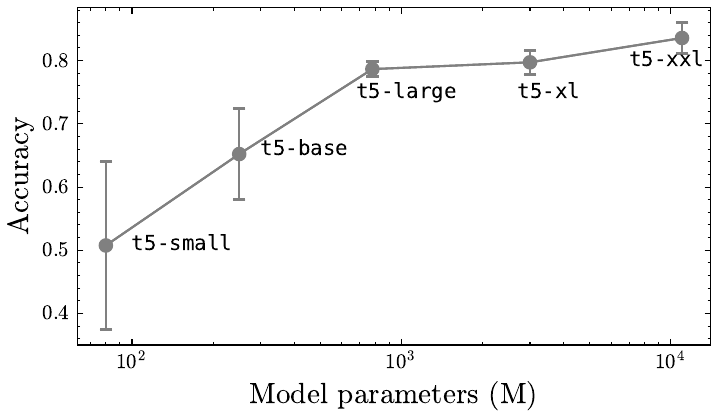}
    \caption{Comparing the matching accuracy among different sizes of instruction finetuned models, accuracy is averaged across domains and datasets therein.}
    \label{fig:compare-LLM-accuracy}
\end{figure}

In this section, we investigate the correlation between larger Large Language Models (LLMs) and enhanced accuracy in schema matching tasks, as observed in related works across various domains (e.g., \cite{chung2022scaling,touvron2023llama,chowdhery2022palm}). To explore this relationship, we conduct an experiment comparing downstream accuracy using FLAN-T5~\cite{chung2022scaling} as the backend with varying LLM sizes (Small-80M, Base-250M, Large-780M, XL-3B, XXL-11B). Evaluation settings, including filter hyperparameters, remain consistent across all assessments. The average accuracy across four domains plotted against model sizes is depicted in Fig.~\ref{fig:compare-LLM-accuracy}.
\subsection{Effect of number of shots to matching accuracy}
\begin{figure}
    \centering
    \includegraphics[width=0.85\linewidth]{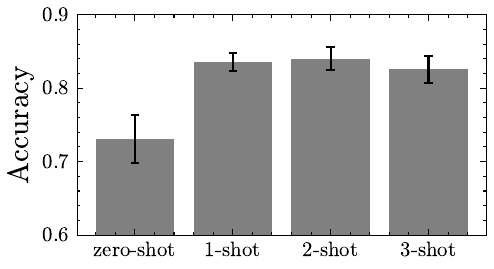}
    \caption{Comparing matching accuracy under varying $k$-shot examples in the prompt.}
    \label{fig:k-shot-experiment}
\end{figure}

In this experiment, we explore the effect of adding more ground-truth examples in the prompt for LLM to conduct in-context learning. It is widely believed that more diverse few-shot examples typically converts to higher accuracy. However, we noticed that the return of adding more samples diminishs very quickly beyond $1$-shot setting. In Fig.~\ref{fig:k-shot-experiment} we showed an significant accuracy boost moving zero-shot (73.05\%) to $1$-shot (83.58\%), whereas the accuracy improvement beyond $1$-shot is not significant given the error band. This finding led us to configure our model to consume just one example per class label.

\subsection{Launch strategy, user experience, and learning}
Due to the service availability requirement, we reserved three nodes on each region with instance type \texttt{ml.g5.2xlarge} as well as several back-up instance types (\texttt{ml.g4dn.4xlarge} etc.) so that in case one instance type isn't available at a certain region we will still be able to serve the request at similar throughput. Another challenge is the volatility of payloads: some payload consists of small schema ($\le 30$ attributes) while in extreme cases this number can be as large as $120$ attributes. To prevent the requests from queuing up, we distribute the attribute mapping requests originating from the same table to at least $3$ hosts with an automatic scaling-up policy.
\par
We present the schema matching service to customers by enabling a human-in-the-loop process: schema matcher never generates the final mapping result in one shot, instead, customers have the chance to examine the predicted mapping table as well as other machine generated metadata (such as searchable keys) and correct any incorrect predictions on the fly. While we are not permitted to record the user activities (e.g. number of modifications they made when composing the schema mapping) due to data privacy, internal studies show that with our LLM aided schema mapping, the amount of human efforts measured by editing operations reduced by 90\%.
\par
Lastly, we also learned from some negative feedback, mostly about the instability of prediction results. Although the model performs reliably on canonical input schema, it predicts wrongly when we slightly change the column name. For instance, by adding a meaningless prefix "XYZ\_" to all column names, the mapping accuracy drops under certain inputs (although not very common). We attribute this as adversarial examples and we plan to focus on this problem as the next research topic.
\section{Discussion}
The schema matching task has been under investigation for over a decade. We posit that the fundamental challenge stems from comprehending attributes in highly heterogeneous environments. The rapid evolution of large language models has elevated language understanding capabilities to unprecedented levels. In light of this advancement, we address the longstanding and intricate problem using this innovative tool, yielding encouraging results. Looking ahead, our future direction involves contemplating the optimal approach for task adaptation to the backbone model, with the aim of further enhancing matching accuracy.

\bibliographystyle{ACM-Reference-Format}
\balance
\bibliography{sample-base}

\appendix

\section{Dichotomies of Schema-matching}
Before delving into the historical overview of schema matching research, it is pertinent to highlight the dichotomies that characterize existing ideas, as elucidated in the review paper by Rahm and Bernstein \cite{rahm2001matching}:

\begin{compactitem}
    \item \textit{Schema-only} or \textit{schema+instances}: A matching system is categorized as schema-only when it relies solely on schema data without considering column values. In contrast, schema + instances matching incorporates both schema and column values. In the context of modern machine learning, the former is often referred to as \textit{zero-shot}.
    \item \textit{Element-wise} or \textit{structural} matching: Element-wise matching entails pairing individual attributes, while structural matching involves matching groups of attributes together.
    \item \textit{Linguistic-based} or \textit{rule-based}: Linguistic-based matching encompasses ideas that employ machine learning or non-machine learning-based text similarity metrics to determine attribute equivalence. Conversely, rule-based matching relies more on schema constraints, such as data types, value ranges, uniqueness, etc.
    \item \textit{One-to-one} or \textit{many-to-many}: A one-to-one matcher consistently connects one attribute to another, whereas a many-to-many matcher has the capability to associate more than one attribute as the source or destination.
    \item \textit{Self-contained} or \textit{auxiliary information}: A self-contained matcher operates autonomously, while a matcher incorporating auxiliary information can leverage external knowledge, such as dictionaries, global schemas, previous matching decisions, and user input.
\end{compactitem}

Having elucidated the aforementioned dichotomies, our focus now shifts to a comprehensive review of both seminal contributions and the current state-of-the-art in the field of schema matching.

\section{Implementation detail of NER filter}
\begin{figure}[htb]
    \centering    \includegraphics[width=0.9\linewidth]{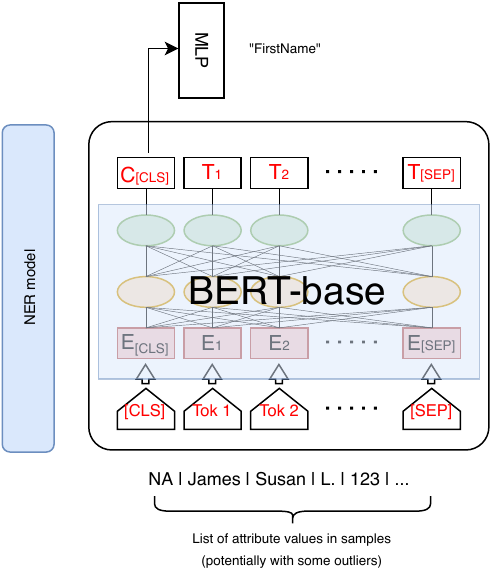}
    \caption{NER model design and input structure.}
    \label{fig:NER-design}
\end{figure}

We follow the identical modelling steps as standard BERT-based NER~\cite{DBLP:journals/corr/abs-1810-04805,tjong-kim-sang-de-meulder-2003-introduction}. The model architecture (as well as input structure) is illustrated in Fig.~\ref{fig:NER-design}. We highlight that the input sequence to the NER model is not a single attribute value, but a list of example values in variable length $k$ with noises (such as empty values, invalid values, etc). Adding external noise helps robustifying the model inference, as it simulates the outliers often encountered in real applications.

Different from previous design of NER models, here we consider a much more broader and finegrained labels, specifically, we consider following categories:
\begin{compactitem}
\item FirstName: Indicates people's first name.
\item MiddleName: Indicates people's middle name.
\item LastName: Indicates people's last name.
\item FullName: Indicates people's first name + middle name (optional) + last name.
\item BusinessName: Indicates a company name. Such as \emph{Amazon.com inc.}
\item ProductName: Indicates the name (title) of a product. Such as \emph{Apple Iphone 13 pro 128GB}.
\item Dates: Indicates a date string in any format compliant to ISO8601. Such as 1989-02-27.
\item Gender: Indicates people's gender identities.
\item Email: Indicates a valid email address, such as \emph{xyz@gmail.com}.
\item URL: Indicates a valid URL, such as \emph{https://www.google.com}
\item CreditCardNumber: Indicates a credit card number string.
\item Timestamps: Indicates a full datetime in at least seconds. Such as \texttt{2001-03-14T19:43:01.342998}.
\item AddressLine: Indicates address line 1.
\item City:  Indicates a city name.
\item Province/State: Indicates a province or state name.
\item Counties: Indicates a country name.
\item Zip/PostalCode: Indicates a zip code.
\item PhoneNumber: Indicates a phone/mobile number with optional area code.
\item Prices: Indicates product prices, such as $12.29\$$.
\item Currencies: Indicates currency symbol, such as \$, JPY, CAD, etc.
\item Weights/units: Indicates the weight or unit of products, such as $2$lbs, $15$ct.
\item FreeText: The fall-back category not captured by any of the above labels.
\end{compactitem}

A majority of data categories can be synthesized by random generation. Part of the data are collected from the internet / open-source datasets; while we also collected some useful examples with LLM prompting, similar to the idea of UniversalNER~\cite{zhou2023universalner}. In total we have 10,000 data entries. During training, we leverage the idea of mixup~\cite{zhang2017mixup} to further augment the training dataset, in case there are multiple different categories in the input, we also create soft-labels when computing the training loss. 
\end{document}